\documentstyle[11pt,twoside,jltp]{article}

\title{Baryon  Asymmetry of Universe: View from Superfluid  $^3$He}

\author{G.E. Volovik\address{Helsinki University of Technology,  Low Temperature
Laboratory, P.O.Box 2200, FIN-02015 HUT, Finland} \address{permanent address:
Landau Institute for Theoretical Physics, Moscow, Russia}}

\runninghead{G.E. Volovik}{Baryon  Asymmetry of Universe:  View from
Superfluid  $^3$He}

\begin{document}

\begin{abstract}
The origin of the excess of matter over antimatter  in our
Universe remains one of the fundamental problems. Dynamical baryogenesis in the
process of the broken symmetry electroweak transition in the expanding Universe
is the widely discussed model where the baryonic asymmetry is induced by
the quantum chiral anomaly. We discuss the modelling of this
phenomenon in superfluid $^3$He and superconductors where the chiral
anomaly is realized in the presence of quantized vortex, which
introduces nodes into the energy  spectrum of the fermionic
quasiparticles. The spectral flow of fermions through the
nodes during the vortex motion leads to the creation of
fermionic charge from the superfluid vacuum and to
transfer of the  superfluid linear momentum
into the heat bath, thus producing an extra force on the
vortex, which in some cases compensates the Magnus force. This
spectral-flow force was calculated 20 years ago by Kopnin and
Kravtsov for $s$-wave superconductors,  but only recently was it
measured  in a broad temperature range in Manchester experiments
on rotating superfluid $^3$He.  The "momentogenesis" observed in $^3$He
is analogous to the dynamical production of baryons by
cosmic strings. Some other possible scenaria of
baryogenesis related to superfluid $^3$He are
discussed.

\end{abstract}

\maketitle


\section{INTRODUCTION}
The superfluid phases of $^3$He and the electroweak vacuum
of the standard Weinberg-Salam model  share many common properties
\cite{VolovikVachaspati} (see Fig.1). The most important similarity is that the
fermionic quasiparticles in superfluid $^3$He interact with the order parameter
in a way that is similar to the interaction of fermions with the electroweak
gauge and Higgs fields. Within this analogy the baryon number violation  appears
to be analogous to the violation of fermion momentum conservation within an
A-phase vortex. The phenomenon which unites the barygenesis in the
Weinberg-Salam model (electroweak baryogenesis) and the momentogenesis in the
pair-correlated fermionic systems is known generically as the ``chiral anomaly".
There are certain quantities, of which baryon number is an example, which are
classically conserved but can be violated by quantum mechanical effects. The
process leading to particle creation is called ``spectral flow'', and can be
pictured as a process in which fermions flow under an external perturbation from
negative energy levels towards positive energy levels.  Some fermions therefore
cross zero energy and move from the Dirac sea (vacuum) into the observable
positive energy world (matter), introducing an extra charge of the matter.

The analogous process in superfluids/superconductors is the transfer of the
linear (or angular) momentum from the condensate (vacuum) to the heat bath
of the normal excitations (matter). Thus the momentum of the ``matter''
is not conserved, while the conservation of the total momentum of vacuum+matter
is not violated. The striking feature of this analogy is that the rate of
momentum transfer due to the spectral flow is described by the same equation of
the chiral anomaly which was introduced  by Adler \cite{Adler1969} and Bell and
Jackiw\cite{BellJackiw1969} in quantum field theory. This is vusualized on the
basis of the vortex texture in $^3$He-A. However, such anomalous violation of
fermionic charge is a general phenomenon in condensed matter and
occurs also for vortices in $^3$He-B and Abrikosov vortices in superconductors.

\subsection{Chiral Anomaly.}
It is known that the charge  particle with definite chirality, say, right-handed
electron, has a peculiar property in the presence  of magnetic field.  The
Landau
quantization of the motion of this particle in magnetic field ${\bf B}\parallel
\hat  z$ leads to an anomalous gapless branch
$E(p_z)= cp_z$, which crosses zero energy at $p_z=0$ (Fig.2). This branch is
asymmetric with respect to the spatial inversion $p_z \rightarrow - p_z$. As a
result the vacuum state is asymmetric since only the states with negative $p_z$
are occupied.  The spectral flow of the electrons from the vacuum state to the
positive energy world leads to the nucleation of the matter from the vacuum.
This occurs if the electric field is applied along $z$. According to the
second Newton law the linear momentum of the particle increases linearly with
time, $\partial_t p_z = e_R E_z$, where $e_R$ is the electric charge of right
electron. This means that under applied field the particles from negative energy
levels of Dirac vacuum  are pushed to positive energy levels of matter with
the rate determined by the electric field projection $E_z$ on the magnetic field
and by the density of state $|e_R{\bf B}|/(2\pi)^2$ on the lowest Landau level:
$$\partial_t n = {1\over   4\pi^2} e_R^2 {\bf E}\cdot {\bf B}
\eqno(2.1.1a)
$$

The left-handed electrons have the anomalous branch with the opposite slope
$E(p_z)=- cp_z$, as a result the production of the particles from the vacuum
has the opposite sign:
$$\partial_t n = - {1\over   4\pi^2} e_L^2 {\bf E}\cdot {\bf B}
\eqno(2.1.1b)
$$
where $e_L$ is the charge of the left electron. If the parity is conserved,
then $e_L=e_R$, and the  net
production of the particles from the vacuum is zero. In
electroweak interactions and in
$^3$He-A the parity is broken, and this leads to the chiral anomaly:
uncompensated anomalous production of the charge from the vacuum.

\subsection{Baryon Production by Chiral Anomaly.}

In the Weinberg-Salam model there are two types of magnetic and electric
fields: hypercharge fields ${\bf B}_Y$ and ${\bf E}_{Y}$ and weak fields ${\bf
B}_W$ and ${\bf E}_{W}$. The   hypercharge
$Y$,  and weak charge $W$ of left and right
quarks are essentially different:
$$Y_{dR}= -{1\over 3} ,~ Y_{uR}= {2\over 3} ,~
W_{dL}=-W_{uL}= {1\over 2}  ,~  W_{dR}=W_{uR}= 0  ,~   Y_{dL}=Y_{uL}={1\over
6}
\eqno(2.2.1)$$
Here the subscripts $uL$ and $uR$ mean  the left and right $u$-quarks; the same
is for the $d$-quarks. The total number of the $u$- and $d$- quarks produced
by the electric and magnetic fields per unit time per unit volume is thus
$$  {3\over 4\pi^2}[{\bf B}_Y\cdot {\bf E}_{Y}
 (Y_{dR}^2 +Y_{uR}^2 - Y_{dL}^2 - Y_{uL}^2) +  {\bf B}_W\cdot {\bf E}_{W}
 (W_{dR}^2 +W_{uR}^2 - W_{dL}^2 - W_{uL}^2)] ~~,
\eqno(2.2.2)$$
where the factor 3 appears due to 3 colours of quarks.
Substituting all the charges from Eq.(2.2.1) and taking into account that there
are $N_F=3$  families of quarks and that all the quarks have the same baryon
number $B=1/3$  one  obtains the following rate of the baryoproduction
$$\partial_t B
= {{N_F} \over {8 \pi^2}} \left (
-  {\bf B}_W\cdot {\bf E}_{W} +
   {\bf B}_Y\cdot {\bf E}_{Y} \right)~.
\eqno(2.2.3)
$$
The nonconservation of $B$ is
thus the result of the phenomenon of the chiral (axial) anomaly predicted by
Adler \cite{Adler1969} and Bell and Jackiw\cite{BellJackiw1969}. The fields
${\bf B}_Y$, ${\bf E}_{Y}$, ${\bf B}_W$ and ${\bf E}_{W}$ can
be generated in the core of topological defects, such as monopoles, domain
walls, sphalerons and electroweak cosmic strings evolving in the expanding
Universe
\cite{Dolgov,Vilenkin,HindmarshKibble,Turok,tvgf,jgtv}.

\subsection{Momentogenesis in 3He-A.}

The anomaly mediated transfer of the linear
momentum can be visualized through the example of the continuous distribution of
the vector
${\hat {\bf l}}$ in
$^3$He-A, which shows the direction in the momentum space, where the
spectrum of quasiparticles has a node. In the vicinity of the point gap
nodes the quasiparticles  are chiral: like
the neutrino they are either left-handed or right-handed
\cite{Exotic}. The  interaction of the quasiparticles with the ${\hat{\bf l}}$
texture is characterized by the ``vector potential'' ${\bf A} =k_F{\hat {\bf
l}}$, where  $p_F=\hbar k_F$ is the magnitude of the momentum at the position of
the node. For such gapless fermions the anomaly in Eq.(2.1.1) takes place
producing a ``chiral charge'' of quasiparticles with the rate
$ (1/ 2\pi^2)~{\bf E} \cdot
{\bf B} =(k_F^2/ 2\pi^2)
~\partial_t {\hat {\bf
l}} \cdot (\vec \nabla \times {\hat {\bf
l}}) $ where the {\it electric} and {\it
magnetic} fields are ${\bf E}=\partial_t {\bf A}$, ${\bf
B}=\vec\nabla\times {\bf
A}$.
The linear momentum produced by the chiral anomaly is
$$
\partial_t {\bf P} = {1\over {4\pi^2}} ({\bf B}\cdot {\bf E}) \left[ {\bf
P}_R  -
{\bf P}_L \right]~~. \eqno(2.3.1)
$$
where
$ {\bf P}_R$   and  $ {\bf P}_L$  -- momenta of  right and left quasiparticle.
The Left-Right  asymmetry,  $ {\bf P}_R=- {\bf P}_L=p_F{\hat {\bf l}} $, gives
the net momentoproduction
$$
\partial_t {\bf P}=
{1\over {2\pi^2}}\int d^3r~ p_F\hat {\bf l} ~({\bf B}\cdot {\bf E})=
\hbar{k_F^3\over {2\pi^2}}\int d^3r~  \hat {\bf l} ~(\partial_t \hat {\bf l}
\cdot (\vec \nabla \times \hat {\bf l} \, \, ))~~. \eqno(2.3.2)
$$
Since the total linear momentum is conserved in condensed matter, the
Eq.(2.3.2)  means that, in the presence of a
time-dependent texture, the momentum is transferred from the superfluid {\it
vacuum} to the {\it matter} (the system of the excitations).

\subsection{Momentogenesis by a Continuous Vortex}

The simplest time-dependent texture, which is experimentally realized in
$^3$He-A, is the moving continuous vortex.  The latter corresponds to
topologically unstable continuous strings in electroweak theory
\cite{VolovikVachaspati}. The simplest type of continuous vortex has the
following distribution of
${\hat{\bf l}}$-field and  superfluid velocity ${\bf v}_s$ in cylindrical
coordinate system:
$$
{\hat{\bf l}}(r,\phi)={\hat{\bf z}} \cos\eta(r) + {\hat{\bf
r}} \sin\eta(r)~,~{\bf v}_s(r,\phi)=-{\hbar\over 2 m r}[1+\cos\eta(r)]{\hat
{\bf \phi}}~~,
\eqno(2.4.1)
$$
where $\eta(r)$ changes from $\eta(0)=0$ to $\eta(\infty)=\pi$ in the so called
soft core of the vortex. The stationary vortex generates the
``magnetic'' field
${\bf B}=k_F\vec \nabla \times {\hat {\bf l}}$. When the vortex moves with a
constant velocity ${\bf v}_L$ (with respect to the heat bath) it also generates
the ``electric''  field, since ${\bf A}$  depends  on
${\bf r}-{\bf v}_Lt$:
$${\bf E}=\partial_t {\bf A}=-k_F({\bf v}_L\cdot
\vec\nabla){\hat {\bf l}} ~~. \eqno(2.4.2)
$$
Integration of the anomalous momentum transfer in Eq.(2.3.2) over the
cross-section of the soft core of the moving vortex gives the rate of the
momentum transfer between the condensate and the heat bath,
mediated by the moving vortex \cite{Volovik1992}:
 $$ \partial_t {\bf P}=-\pi \hbar N_v  C_0{\hat {\bf z}}  \times {\bf
v}_L ~~.
\eqno(2.4.3)
$$
Here $N_v$ is the winding number of the vortex ($N_v=-2$ in the case of
Eq.(2.4.1)) and $C_0=  k_F^3/3\pi^2$.

This corresponds to an extra nondissipative force $ {\bf F}_{sf}=\partial_t {\bf
P}$ acting on the moving continuous vortex, which is called the spectral-flow
force. This result for the continuous vortex, derived from
the axial anomaly equation (2.1.1), was confirmed by the microscopic theory,
which took into account the discreteness of the quasiparticle spectrum in the
soft core \cite{Kopnin1993}.

\subsection{Spectral-flow force vs Magnus force.}

The spectral-flow force is the transverse nondissipative force which acts on the
vortex, when it moves with respect to the matter (heat bath or normal
component):
$$ {\bf F}_{sf}=-\pi \hbar N_v  C_0{\hat {\bf z}}  \times [{\bf
v}_L-{\bf
v}_n] ~~,
\eqno(2.5.1)
$$
where ${\bf
v}_n$ is the velocity of the normal component of the liquid.
In addition to this there is also an old classical Magnus force, which acts
on the vortex moving  with respect to superflow:
$$ {\bf F}_{M}= \pi \hbar N_v  {\rho\over m} {\hat {\bf z}}  \times [{\bf
v}_L-{\bf
v}_s] ~~,
\eqno(2.5.2)
$$
where ${\bf
v}_s$ is the superfluid velocity of the coherent condensate (vacuum) and
$\rho/m$
is the particle density.
The balance of these two forces acting on the vortex (together with the
Iordanskii force, which we do not discuss here, see Ref.\cite{Sonin}, and  the
longitudinal friction force, see below in Sec.~3.2) determines the motion of the
vortex and thus the mutual force between the vacuum and matter mediated by the
moving vortex.

It is important, that if   ${\bf v}_n={\bf
v}_s$ the spectral flow almost completely compensates the  Magnus force. This is
because the particle density $\rho/m$ and the spectral-flow parameter $C_0=
k_F^3/3\pi^2$ are very close to each other. In fact in the normal Fermi liquid
above the superfluid transition temperature $T_c$ these quantities coincide:
$\rho(T>T_c) =  m k_F^3/3\pi^2$. In the superfluid phase the disbalance between
$\rho/m$ and  $C_0$ occurs due to the tiny asymmetry between particles and
holes: $\rho(T<T_c) -  m k_F^3/3\pi^2 \sim \rho (\Delta(T)/E_F)^2 \ll \rho $,
where
$\Delta(T) $ is the  gap amplitude and $E_F$ the Fermi-energy.
Such cancellation of the Magnus force was  confirmed in  experiments on the
vortex dynamics in $^3$He-A \cite{BevanNature,BevanNew}.

\section{MOMENTOGENESIS BY SINGULAR VORTICES}

The momentum is transferred from the vacuum to the heat bath
by the vortex motion, the chiral anomaly being the main mechanism which
determines the rate of momentum transfer. However, in the case of
vortices with a singular core (in superconductors and in $^3$He-B) the equation
(2.1.1) for the axial anomaly is no more valid: the anomalous
force should be  calculated directly from the consideration of the spectral flow
in the  vortex core. This is similar to the calculation of the baryonic
charge on
$Z$-strings, where instead of the anomaly equations one must operate  with the
exact spectrum of fermions  in the string core \cite{jgtv}.

\subsection{Spectral Flow in Singular Vortex.}

The spectrum of single-fermion excitations has anomalous
(chiral) branch $E(p_z,L_z)$, which  depends on the momentum
projection $p_z$ on the vortex axis and the discrete angular momentum $L_z$
in the following way \cite{Kopnin1993}:
$$
E(p_z,L_z)=- \omega_0(p_z)~(L_z-L_{0}(p_z)) ~~ \eqno(3.1.1)
$$
For most symmetric vortices, where $L_{0}(p_z)=0$, the spectrum was first
obtained in \cite{Caroli}.
The interlevel distance $\hbar\omega_0(p_z)$, the so called minigap, is small
compared to the gap amplitude $\Delta(T)$ of the fermions  outside the
core:
$\hbar\omega_0\sim \Delta^2(T)/E_F \ll \Delta(T)$, where again $E_F$ is the
Fermi-energy. As a
function of $L_z$, the anomalous branch  crosses zero  energy (Fig.3). According
to the index theorem, the number of such anomalous branches equals the
winding number
$N_v$ of the vortex \cite{Volovik1993}). This results in a  momentum flow
from the the Dirac sea to the heat bath, if the
minigap
$\hbar\omega_0$ between the core states can be neglected. We consider just this
case, which occurs when $\omega_0$ is less than the inverse relaxation time,
$1/\tau$.

If the velocity $v_L$ is along
$x$, the  excitation angular momentum $L_z$ in the moving
core changes with the rate
$ \dot L_z=\dot x p_y=v_L p_y$. Since the neighbouring levels differ by
$\Delta L_z=\hbar$ and  the number of anomalous branches equals the winding
number $N_v$, the levels cross zero at a rate $-N_v\dot L_z/\hbar$. Thus the
flow of the momentum $p_y$ from the negative to the positive energy
states occurs with the rate $\dot p_y=-N_v p_y
\dot L_z/\hbar= - N_v v_L p_y^2/\hbar$. Integrating this momentum transfer over
the Fermi surface and taking into account that the in-plane average $<p_y^2>
={1\over 2} (p_F^2-p_z^2)$, one again obtains  the Eq.(2.4.3)
$$
\partial_t{\bf P}={\hat {\bf y}}\partial_t P_y =-{1\over 2 \hbar}{\hat {\bf y}}
v_L  N_v\int_{-p_F}^{p_F} {dp_z\over 2\pi\hbar} (p_F^2-p_z^2)=-  N_v \pi \hbar
C_0 {\hat {\bf z}}
\times {\bf v}_L ~~. \eqno(3.1.2)
$$

Thus the result of the direct calculation of
the spectral flow in the singular core coincides with the  result obtained for
the continuous $^3$He-A vortex from the axial anomaly equation. This is not a
coincidence, but the consequence of the underlying topology in the 6-dimensional
real+momentum space. The singular vortex in any Fermi superfluid/superconductor
can be transformed into a continuous one by extending the core. The order
parameter field in the extended core is similar to that of  $^3$He-A: the
spectrum of quasiparticles  contains the point nodes\cite{VolovikMineev1982}. In
the vicinity of the nodes the quasiparticles are chiral leading to the
Adler-Bell-Jackiw anomaly and thus to the spectral-flow force.

\subsection{Nondissipative and Friction Forces.}

The anomalous nondissipative force acting on the vortex has maximal value if the
spectral flow is not suppressed by the minigap, i.e. at
$\omega_0\tau \ll 1$.  In the
general case considered in
\cite{Stone,KopninVolovik1995} and experimentally  measured in superfluid
$^3$He-B \cite{BevanNature,BevanNew}, the spectral flow
force is reduced by the factor $1/(1+\omega_0^2\tau^2)$ and also the dissipative
friction component of the force appears:
$$
{\bf F}_{sf} =   N_v \pi \hbar C_0 \left[{1\over 1+\omega_0^2\tau^2} {\hat
{\bf z}}
\times ({\bf v}_n -{\bf v}_L) + {\omega_0\tau\over 1+\omega_0^2\tau^2}  ({\bf
v}_n -{\bf v}_L)\right]~~, \eqno(3.2.1)
$$
where ${\bf v}_n$ is the velocity of the heat bath (or velocity of the normal
component of the superfluid/superconductor).   This expression was
found more than 20 years ago by Kopnin and Kravtsov in the Gorkov formalism
developed for
$s$-wave superconductivity
\cite{KopninKravtsov}. In the $d$-wave
superconductor the situation is more interesting due to the gapless
quasiparticles outside the vortex cores
\cite{KopninVolovik1997a}. In this case the minigap $\omega_0$ depends on the
orientation and has nodes. This leads to the finite dissipation even in the
superclean limit $<\omega_0>\tau =\infty$, which occurs due to the Landau
damping. In the moderately clean regime the
scaling law for the flux flow at low $T$ and low field is obtained. The gapless
regime can also occur for $^3$He-B vortices with the spontaneously broken
continuous symmetry  of the core, this leads to effects corresponding to the
Hawking radiation at the event horizon in black holes
\cite{KopninVolovik1997b,JacobsonVolovik}.

\section{TOPOLOGY OF MOMENTOGENESIS}

\subsection{Topological and Fermionic Charges}

The vortex is the object which serves as a mediator between the
condensate (vacuum) and heat bath (matter) and  which allows them to exchange
linear momentum, while the total momentum is conserved:
$${\bf P}={\bf P}_{vac}+{\bf P}_{matter}=\int d^3r~
 \rho   {\bf v}_s + \sum_{\bf p} f({\bf p}){\bf p}~~.\eqno(4.1.1)$$
Here ${\bf v}_s$ is the superfluid velocity which in terms of the phase $\Phi$
of the order parameter
is ${\bf v}_s={\kappa\over 2\pi}\vec \nabla \Phi$;
$\kappa$  is the
circulation quantum,  $\kappa=\pi\hbar/m_3$  for the   superfluid
$^3$He, with $m_3$ being is the mass
of $^3$He atom); $\rho$ is the mass density and
$f({\bf p})$ is the distribution function for the quasiparticles.  The dynamics
of the momentogenesis is determined by the spectral-flow.

Another property of
the vortex is that it can change the topological charge of the
vacuum. The vortex thus serves as the instanton, if the
transition between the vacua of different topological charges occurs by
tunneling, or as the sphaleron, if the transition occurs by thermal activation.
In the Weinberg-Salam model the transition between topologically different vacua
is accompanied by a change of the baryonic charge. The same happens with the
fermionic charge of the superfluid.
In the torus geometry the relevant  fermionic charge of the vacuum is the
momentum
$L_z$, while the topological charge is the winding number of the phase
$\Phi$ along the channel:
$$L_z(vac)= {\hat {\bf z}}\cdot \int d^3r~ {\bf r}\times   (\rho {\bf v}_s)
~~,~~N(vac)={1\over \kappa}\oint d{\bf r}\cdot{\bf v}_s~~.\eqno(4.1.2)$$
These two charges are related:
$${L_z(vac)\over \hbar}= {1\over 2}{\cal N}(vac)   N(vac)~~,~~{\cal
N}(vac)={\rho
\over m_3} V_{\rm torus}~.\eqno(4.1.3)$$
Here ${\cal N}(vac) $ is the total number of the $^3$He atoms
in the vacuum within the torus volume
$ V_{\rm torus}$.
Note that $  L_z(vac)/\hbar$
is always integral, even if the total number of $^3$He atoms is odd. In the
latter case one atom has no partner to form a Cooper pair  and thus is not
in the vacuum even at $T=0$: it belongs to the {\it matter}. Another example
of the nonzero fermionic charge of the {\it matter} even in the ground state
will be discussed in Sec.5.1.

The change of the topological charge of the vacuum
occurs by nucleation of a segment  of the
vortex with winding number $N_v$  on the surface of the channel. The vortex
sweeps the cross section of the channel and then disappears on the surface.
In this process  the  topological charge
$N(vac)$  changes by $N_v$. The superflow velocity in the channe,
$v_s=\kappa N(vac)/  L$, decreases by the value $\Delta v_s=\kappa N_v/  L$
(where $L$ is the length of the annular channel) which leads to the change
of the fermionic charge of the vacuum  by
$ \Delta L_z(vac)/ \hbar ={1\over 2} {\cal N}(vac)   N_v$.
In electroweak theory this  corresponds to the relation between
the change of topological number of the vacuum (Chern-Simons number $N_{CS}$)
and the change of the baryonic charge $\Delta B$:
$$
\Delta B =  N_F \Delta N_{CS}\ ,
\eqno (4.1.5)
$$
where $N_F$ is the number of the fermionic families. The difference from
condensed matter case is that in the relativistic theories with local gauge
symmetry the energy of the vacuum does not depend on its topological quantum
number, while in the case of superfluid/superconductor the energy of the vacuum
is
$\propto v_s^2\propto N^2(vac)$.

\subsection{Quantum nucleation. Instanton.}

The instanton is the
quantum tunneling between two vacua, which is realized by the quantum
nucleation of the vortex loop.  At low
$T$ the spectral flow is suppressed and the tunneling is determined by the
classical action which kinetic term is related to the Magnus force:
$$ S= \kappa N_v \rho V +\int dt~E~~.\eqno(4.2.1)$$
Here $E$ is the energy of the vortex line. The first term is the kinetic part
and is an analog of the Wess-Zumino action, where $V$ is the volume bounded by
the area swept by the vortex loop between nucleation and
annihilation\cite{Rasetti}.  Variation of this term leads to the Magnus
force. This term is of topological origin and leads to  quantization of the
total number number of particles  in the
vacuum, ${\cal N}(vac)=(\rho/m) V_{total}$,  where
$V_{total}$ is the total volume of the vessel. The quantization occurs due to
the ambiguity in choosing the volume swept by the loop: it can be  $V$ or
$V-V_{total}$.  Since the exponent
$e^{iS/\hbar}$ should not depend on this choice, the difference
between the actions,
$\Delta S =m\kappa N_v {\cal N}(vac)$, must be multiple of
$2\pi\hbar$. For $^4$He, where $\kappa=2\pi \hbar/m$ this gives an integral
value of
${\cal N}(vac)$, while for $^3$He, where $ \kappa=\pi \hbar/m$, the number
of particles in the vacuum must be even. The volume law for the
action is the property of the global vortices where the field
${\bf v}_s$ generated by the vortex line is not screened by the gauge field.

In quantum tunnelling the action along the instanton trajectory contains an
imaginary part, which gives the nucleation rate
 $\Gamma\propto e^{-2{\rm Im} S/ \hbar}$. Let us consider the large
$N(vac)$ case, when the velocity in the annular channel, $v_s=\kappa
N(vac)/  L$,
is large and the nucleation occurs close to the pinning center at the wall of
container. For the hemispherical pinning center the instanton is the trajectory
of the vortex half-ring coaxial with the pinning center and oriented
perpendicular to the flow (axis $x$). Far from the pinning center the energy of
the vortex loop of radius
$r$  is
$$E=E_0(r) + {\bf
v}_s \cdot {\bf p}(r) ~,~E_0(r) = {1 \over 2} \rho  N_v^2\kappa^2
r  \ln { r\over \xi}~,~ p_x(r)= \pi  \rho  N_v \kappa r^2 ~\eqno(4.2.2)$$
where $E_0(r)$
is the line tension, $\xi$ is the core radius, and ${\bf
p}(r)={\hat{\bf x}} p_x$ is the linear momentum of the loop.
The vortex nucleated from the vacuum state must have zero energy, $E=0$. This
determines the radius of the  nucleated vortex:
$$R_0={\kappa N_v\over 2\pi v_s}\ln {R_0\over
\xi}~~.\eqno(4.2.3)$$
The calculation of the instanton action for the $^4$He vortex with
$N_v=1$ shows that it is proportional to the number ${\cal N}_0$ of
particles involved in the process of tunneling:\cite{Volovik1972}
$$ {\rm Im} S  = 2\pi \hbar{\cal N}_0 ~~,~~{\cal N}_0={\rho\over m}
{2\pi\over 3} R_0^3~.\eqno(4.2.4)$$
For the $^3$He case and for the vortex tunneling in superconductors, the
situation can be different because of the fermion zero modes in the vortex core,
which lead to the spectral flow. This question is still open.

\subsection{Thermal activation. Sphaleron}

The thermal activation, which can occur  at nonzero $T$, is
determined by the  so called sphaleron\cite{Turok} -- the  saddle-point
stationary solution, which is intermediate between  vacuum states with
different topological number
$N(vac)$. In  superfluids the sphaleron is
represented by a metastable vortex loop, stationary in the
heat-bath frame: its velocity
$dE/d{\bf p}=0$, which corresponds to the maximum of the Doppler shifted energy
$E=E_0(r) + {\bf
v}_s \cdot {\bf p}(r)$. Using the Eq.(4.2.2) for
$E$ one obtains the radius and the energy  of the sphaleron
$$r_\circ={\kappa N_v\over 4\pi v_s}\ln {r_\circ\over
\xi}~,~E_0 (r_\circ)= {1 \over 2} \rho \kappa^2
N_v^2 r_\circ \ln {(r_\circ/\xi)}
~.\eqno(4.3.1)$$
The sphaleron is created by  thermal activation with a
rate $\exp { -(E_0(r_\circ)/T)}$. After that the vortex ring grows
spontaneously, thus decreasing  its Doppler shifted energy $E$,
until it disappears at the wall of the torus leaving behind the flow pattern
with the topological number
$N(vac)$ reduced by $N_v$.

\section{DISCUSSION. OTHER SCENARIA.}

The quantum field theory in condensed matter and in superfluid $^3$He in
particular can provide analogies for many different scenaria of
baryogenesis. For example, the production of the linear momentum \cite{Yip} and
spin momentum \cite{SchopohlWaxman} during the motion of the interface between
the $^3$He-A and $^3$He-B can simulate the mechanisms related to the
baryogenesis by domain walls.  Here we briefly mention the analogy of
baryogenesis by the baryonic charge condensate \cite{Dolgov}.

\subsection{Nonzero charge of {\it matter} induced by vacuum charge.}

The vacuum with nonzero fermionic charge takes place in ferromagnets, where the
charge is played by the discrete projection $S_z$ of spin to the direction of
the spontaneous magnetization. This charge is  conserved above and below the
broken symmetry phase transition.  Since we are interested in the quantum field
theory, let us consider the hypothetical phases of the superfluid $^3$He,
which have spontaneous vacuum spin $S_z(vac)$. This can be for
example A$_1$-phase or
$\beta$ phase ((Eq.(6.53) in \cite{VollhardtWolfle}), which could appear if the
parameters of the systems were more favourable. The magnitude of the vacuum spin
in such phases is
$$S_z(vac) \sim  \hbar {\cal N}(vac){1\over (k_F \xi)^2} ~~,
\eqno(5.1.1)$$
where ${\cal N}(vac)$ is the total number of the $^3$He atoms;
$\xi=p_F/m\Delta$ is the superfluid coherence length, and $\Delta$ is the gap.
If there are no boundaries and thus no exchange with environment, then
$S_z(vac)$ should be compensated by the opposite amount of the
spin of the quasiparticles. In this case the ground state is not the vacuum: it
contains quasiparticles,  with number
$$N_{qp}=  2 {S_z(vac)
\over \hbar} \sim {\cal N}(vac){1\over (k_F \xi)^2}~~.\eqno(5.1.2)$$
Since $k_F\xi \sim 10^2-10^3$,  the number of the particles in the
{\it matter}, $N_{qp}$, is much smaller than the number of the particles in the
vacuum, ${\cal N}(vac)$, and thus the vacuum is not  disturbed by the {\it
matter}. The energy of the {\it matter} is about $\Delta N_{qp}$ (or even less
if the quasiparticles are massless), and this is essentially smaller than the
vacuum energy.

The  conservation of the total spin after the transition can be also provided by
the formation of two domains with  opposite direction  of  spin. This also
costs energy: now it is the energy of the wall between two domains
$E_{wall} \sim m_3 p_F  \Delta^2 \xi V_{total}^{2/3}$. Comparing this to the
energy of the quasiparticles,  one finds that quasiparticles in the ground
states are preferred over a domain wall if the size of the sample  is
small enough. The condition for the total number of
$^3$He atoms (the total number of the particles in the vacuum) is:
$${\cal N}(vac) <\left( \xi k_F \right)^6 ~~.
\eqno(5.1.3)$$

The symmetry breaking in the Universe can also give rise to
nonzero vacuum fermionic charge, the baryonic number $B(vac)$ \cite{Dolgov}.
Then the matter acquires a baryonic charge opposite to the vacuum charge,
$B_{matter}= -B(vac)$,  if (1) the total baryonic charge is conserved; (2) if
the baryonic exchange between the vacuum and matter is not suppressed; and (3)
if the energy of the domain wall between matter and antimatter is larger than
the energy of the excess baryonic matter.
Many details of this scenario of  baryogenesis can be modelled in condensed
matter. The transition to superfluid $^3$He-A in a small droplet of
$^3$He can lead to  quasiparticles in the ground state, because the A-phase
state has a nonzero vacuum value of the orbital momentum. The decay of the
vacuum spin into the spin of the fermions is one of the heavily discussed
topics in the so-called magnetic superfluidity of $^3$He-B, where the coherent
precession of the magnetization takes place with all the signatures of the spin
condensate \cite{Bunkov}; and so on.

The related problem is the formation of the galactic magnetic
fields. In a recent paper Joyce and Shaposhnikov \cite{JoyceShaposhnikov}
suggested a new scenario: the galactic magnetic
fields are generated from the primordial hypercharge magnetic fields. The
latter are spontaneously formed  by the right-handed electrons through the
Abelian anomaly. It appears that the process of generation of hypercharge
magnetic  field from the fermionic charge of matter is described by the same
equations as the process of the superflow instability in $^3$He-A, which was
intensively discussed theoretically  and has been
recently investigated experimentally \cite{Experiment}. Both processes are
governed by the chiral anomaly. In $^3$He-A, the motion of the normal
component, which contains the fermionic charge -- the linear
momentum of quasiparticles, is unstable towards the inhomogeneous condensate --
the texture of the $\hat {\bf l}$-vector, which corresponds to the hypercharge
magnetic field.

In conclusion, when the problem of the baryogenesis is finally solved, it
will appear that the key to the problem has been provided by  condensed
matter physics.

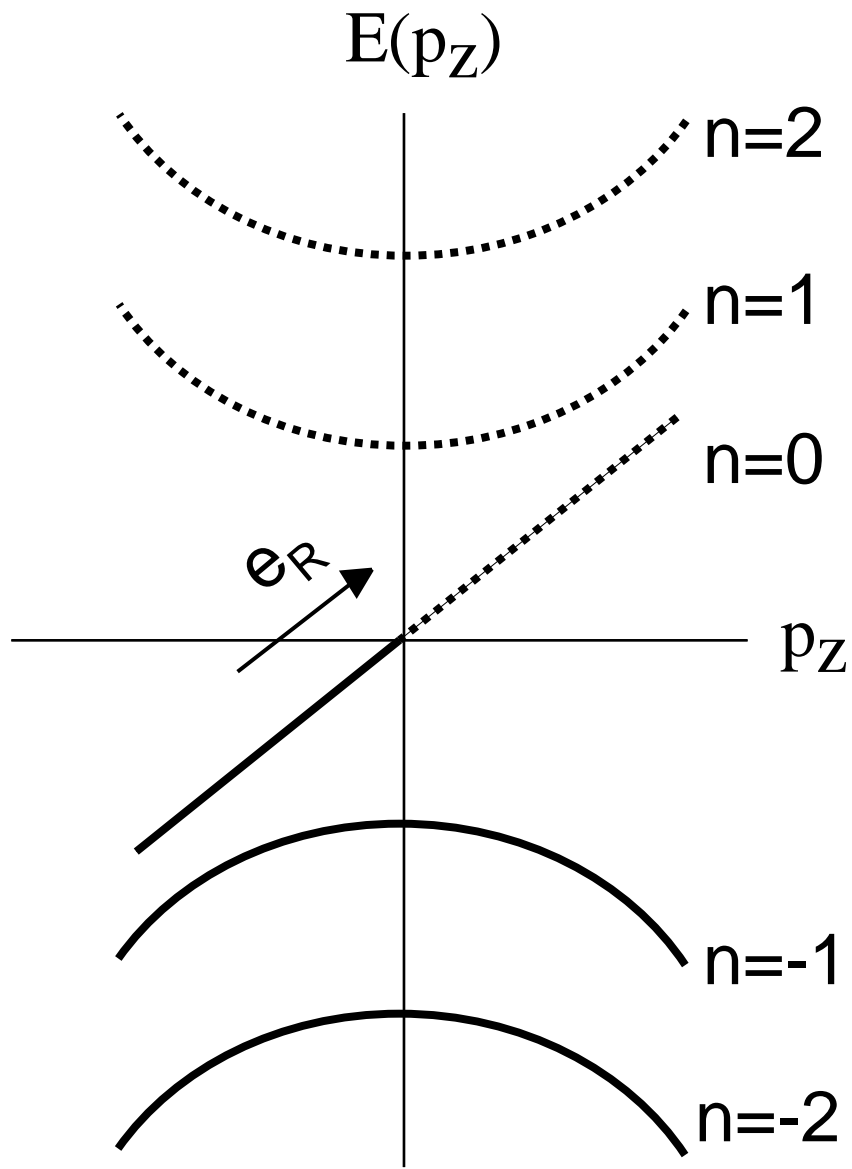
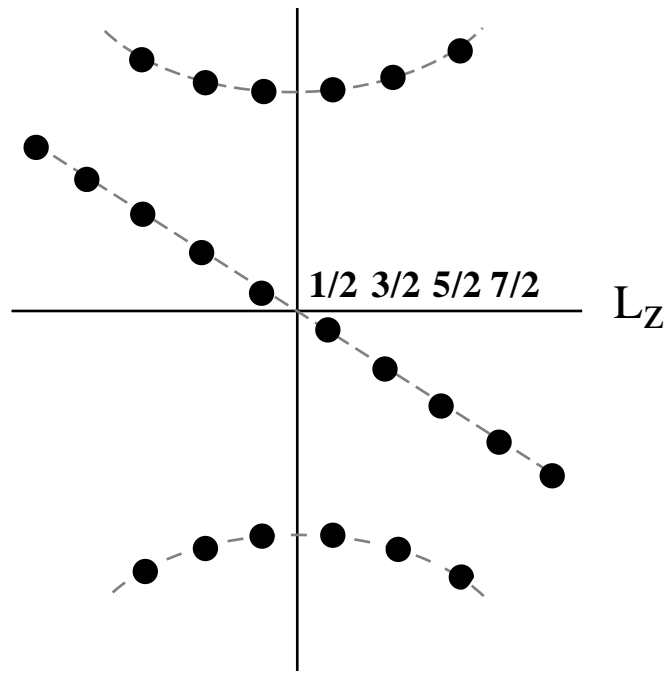
\end{document}